\documentclass[aip,jcp,reprint]{revtex4-2}
\usepackage{amsmath,amssymb,graphicx}

\usepackage{graphicx}
\usepackage{dcolumn}
\usepackage{bm}

\usepackage[utf8]{inputenc}
\usepackage[T1]{fontenc}
\usepackage{mathptmx}
\usepackage{etoolbox}
\usepackage{subfig}
\usepackage{xcolor}
\usepackage{boxhandler}
\usepackage[justification=raggedright,singlelinecheck=false]{caption}
\usepackage{float}

\makeatletter
\def\@email#1#2{%
 \endgroup
 \patchcmd{\titleblock@produce}
  {\frontmatter@RRAPformat}
  {\frontmatter@RRAPformat{\produce@RRAP{*#1\href{mailto:#2}{#2}}}\frontmatter@RRAPformat}
  {}{}
}%
\makeatother
\begin{document}

\preprint{AIP/123-QED}

\title[]{Simulations of dielectric permittivity of  water by Machine Learned Potentials with long-range Coulombic interactions}
 \author{Kehan Cai}
 \thanks{These authors contributed equally to this work.}
\affiliation{ 
 Department of Chemistry, Princeton University, Princeton, New Jersey 08544, USA
 }%
 \author{Chunyi Zhang}
 \thanks{These authors contributed equally to this work.}
  \affiliation{ Department of Chemistry, Princeton University, Princeton, New Jersey 08544, USA
 }%
 \affiliation{ 
Department of Physics, Temple University, Philadelphia, Pennsylvania 19122, USA
}%
\author{Xifan Wu}
\email{xifanwu@temple.edu}
\affiliation{ 
Department of Physics, Temple University, Philadelphia, Pennsylvania 19122, USA
}%
\affiliation{ 
Institute for Computational Molecular Science, Temple University, Philadelphia, Pennsylvania 19122, USA
}%

\date{\today}

\begin{abstract}
    The dielectric permittivity of liquid water is a fundamental property that underlies its distinctive behaviors in numerious physical, biological, and chemical processes. Within a machine learning framework, we present a unified approach to compute the dielectric permittivity of water, systematically incorporating various electric boundary conditions. Our method employs a long-range-inclusive deep potential trained on data from hybrid density functional theory calculations. Dielectric response is evaluated using an auxiliary deep neural network that predicts the centers of maximally localized Wannier functions. We investigate three types of electric boundary conditions—metallic, insulating, and Kirkwood-Fröhlich—to assess their influence on correlated dipole fluctuations and dielectric relaxation dynamics. In particular, we demonstrate a consistent methodology for computing the Kirkwood correlation factor, correlation length, and dielectric permittivity under each boundary condition, where long-range electrostatics play a critical role. This work establishes a robust and generalizable machine-learning framework for modeling the dielectric properties of polar liquids under diverse electrostatic environments.
\end{abstract}

\maketitle

\section{Introduction}

Water is arguably the most important material for life and is involved in almost all biological processes~\cite{eisenberg2005structure}. Despite the apparent simplicity of a water molecule in the gas phase, liquid water in its condensed phase displays several anomalous properties. As one of its unique properties, water has an anomalously large static dielectric constant $\varepsilon \sim 80$ at room temperature~\cite{wyman1930measurements, drake1930measurement, malmberg1956dielectric, sharma2007dipolar}, which makes water unusually sensitive to an external electric field. In response to an applied electric field, water molecules collectively reorientate their electric dipole moments along the field direction. In this way, a large amount of electrical energy is stored, and the external electric field is significantly screened inside the water. The anomalously large dielectric permittivity of water plays a crucial role in numerous physical, chemical, and biological processes. For example, during the dissolving process, the large dielectric constant in water helps to screen the interionic Coulombic interactions, and therefore largely promotes the solubility of salts and polar substances in water. This unique dielectric property contributes to making water referred to as the universal solvent in biology and chemistry, since it can dissolve more substances than any other liquid on Earth~\cite{eisenberg2005structure}. Moreover, the dielectric screening of liquid water also plays a significant role in protein interactions and the formation of electrical double layers in aqueous solutions.

In experiments, the measurement of dielectric constant in water is relatively straightforward, which can be accurately determined by several experimental techniques, such as vacuum tube source of voltage experiment, capacitance-conductance bridge setup, and temperature-controlled microwave cavity methods. One of the first experimental measurements can be dated back to as early as the beginning of twentieth century~\cite{wyman1930measurements, drake1930measurement, malmberg1956dielectric}. In 1930, Drake \emph{et al.} measured the index of refraction in distilled water, and they reported an experimental value of $\varepsilon=78.57$ under ambient conditions~\cite{drake1930measurement}. Later, the dielectric constants in water were extensively studied and its temperature dependence has been measured as well, in which the dielectric constant was found to be decreased with elevated temperatures~\cite{fernandez1997formulation}. Besides the measurement on normal water, the delicate isotope effect has also been accurately determined. Over fifty years ago, a slightly higher dielectric constant in normal water than that in heavy water was reported based on the capacitance experiment~\cite{vidulich1967dielectric}.

In theory, the dielectric constant of water can be calculated based on the knowledge of dipole fluctuation in an equilibrated liquid trajectory within the linear response regime~\cite{sharma2007dipolar}. However, the computer simulations of dielectric constant in water have faced significant challenges over decades. The difficulty lies in the fact that the dipole fluctuation in water is collective in nature which is hard to converge on both time and space scales. In the time domain, a converged dipole fluctuation in water needs to include the oscillation of electric dipoles on water molecules in response to an electric field change, which is characterized by the dielectric relaxation time or referred to as Debye relaxation time. In experiments, the dielectric relaxation time is estimated to be as large as 8-9 picoseconds ($ps$) because of the underlying Hydrogen(H)-bond network. To sufficiently sample the dielectric relaxations in the calculated dipole fluctuations, the liquid trajectories are required to be simulated at a time scale of tens of nanoseconds ($ns$). In the space domain, the theoretical determination of dielectric constant requires to converge the correlated dipole fluctuations on the underlying H-bond network as well.  Because of the directional H-bonding, a water dipole is highly correlated with the dipole on a neighboring water molecule. On a near tetrahedral H-bond network of liquid water, this dipole-dipole correlation decays slowly as the distance between the pair of molecules increases, and a pair of dipoles remain correlated at the scale of nanometers which is characterized by the Kirkwood correlation length. As such, an accurate modeling of dielectric constant demands the modeling of the collective fluctuations of correlated dipoles in real space. In this regard, the liquid water is required to be modeled by large supercell whose size should be comparable to its correlation length. 

To alleviate the computational burden, prior simulations have mostly adopted classical force field based molecular dynamics simulations (MD)~\cite{anderson1987molecular, heinz2001comparison, gereben2011accurate, raabe2011molecular}. Based on empirical models, the predictions from classical force field models usually can be fitted to well reproduce the experimental data. Unfortunately, the MD by classical force fields fails to capture the electric polarizability that gives rise to the significant fluctuations of the H-bonds in liquid water at room temperature. In this regard, {\it ab initio} molecular dynamics (AIMD)~\cite{car1985unified} provides an ideal scheme to simulate the dielectric constant in water. In AIMD, the forces on the nuclei are derived from the instantaneous ground state of the electrons within density functional theory (DFT)~\cite{hohenberg1964inhomogeneous, kohn1965self} while the electrons adjust on the fly and can thereby access the fluctuations of the H-bond network. In equilibrated liquid structures by AIMD, each individual water molecule can be further associated with an electric dipole. Based on modern theory of polarization~\cite{king1993theory}, the electronic contribution to the dipole can be unambiguously assigned to a water molecule from the centers of the maximally localized Wannier functions (MLWFs)~\cite{marzari1997maximally}, which are obtained from DFT eigenstates via a unitary transformation. In a pioneering study, Sharma \emph{et al.} applied AIMD simulation to predict the dielectric constant of water, and the importance correlated dipole fluctuation has been reported~\cite{sharma2007dipolar}. However, the dielectric constant of water has been yet to be accurately determined by AIMD simulations because of the unfordable computational cost as required to converge the dipole fluctuation at both time and space scales.

In the past decade, the advent of artificial intelligence has rapidly pushed this field forward~\cite{behler2007generalized, chmiela2017machine, zhang2022deep, zhang2018deep, zhang2018end}. In the deep potential molecular dynamics approach (DPMD)~\cite{zhang2018deep}, a deep neural network (DNN) is trained by the inputs of AIMD trajectories to represent the many-body interatomic potential as a sum of atomic potentials that depend on the local chemical environment. The deep potential (DP) is extensive by construction, and it can predict the molecular structures of liquid water with the accuracy at DFT level, while its computational efforts are only at the level of empirical force field calculations. In this respect, some initial attempts have been made on {\it ab initio} studies of the dielectric constant of water, in which it was found that the DNN has significantly improved the efficiency in converging the dipole fluctuations in water. Despite the progress, uncertainties remain. So far, most of the simulations adopted the short-range only DP models in the trained DNNs, which implies the neglect of electrostatic interactions at long-range. However, it is known that electrostatic force is long-range in nature~\cite{ballenegger2004structure, cox2020dielectric, gao2022self}; the long-range Coulombic force is essential to correctly predict the asymptotic behavior of dipolar correlation in the classical limit, which is the key underline the determination of Kirkwood correlation length in water. In addition,  {\it ab initio} predictions of dielectric constant of water are further complicated by the treatment on the electric boundary conditions that are applied on the system. The applied electric boundary conditions strongly influence not only the behavior of correlated dipole fluctuation at long range, but also the characteristic in its dynamics in term of the dipole relaxation~\cite{cichocki1989electrostatic}. For example, under metallic and insulating electric boundary conditions, transverse and longitudinal dielectric relaxation times are extracted, respectively; and the calculated correlation factors are distinctly different as well under these two electric boundary conditions. In first-principles calculations, careful analyses should be carried out under various electric boundary conditions before a unified approach can be arrived for the prediction of dielectric constant in water. Unfortunately, such systematic treatment has been elusive so far in the field.

To address the challenges outlined above, we utilize state-of-the-art DNN approaches to model correlated dipole fluctuations in liquid water under ambient conditions and predict its dielectric constant under various electric boundary conditions in a systematic manner. Molecular structures and dynamics are simulated using molecular dynamics based on the long-range inclusive Deep Potential Long-Range (DPLR) model~\cite{zhang2022deep}, with DNNs trained on hybrid density functional theory (DFT) calculations employing the SCAN0 exchange-correlation approximation with $10\%$ exact exchange energy~\cite{zhang2021modeling}. In the DPLR framework, long-range Coulombic interactions among ions and electrons are captured through the electrostatic energy of Gaussian charges positioned at ionic and electronic sites. The electronic sites are defined by the centers of maximally localized Wannier functions, predicted by a second DNN, the Deep Wannier model, trained on electronic structures of water derived from the same DFT calculations~\cite{zhang2020deep}. We explicitly incorporate three common electric boundary conditions—metallic, insulating, and Kirkwood-Fr\"ohlich (KF)—in our simulations. For dynamics, we extract dielectric relaxation times for transverse and longitudinal modes under metallic and insulating boundary conditions, respectively, achieving excellent agreement with experimental data. Furthermore, we demonstrate that the Kirkwood correlation length and correlation factor can be computed across all three boundary conditions, provided the asymptotic behavior of dipolar pair correlations is realized in the classical limit. The long-range Coulombic interactions in DPLR are critical for this purpose. To ensure a converged simulation of the dielectric constant, collective dipole fluctuations must fully capture both dielectric relaxation and correlated dipole fluctuations across temporal and spatial scales. Using this approach, we determine the dielectric constant of water under all electric boundary conditions based on the Kirkwood-Onsager theory.

The remainder of this paper is organized as follows. In Sec. II, we describe the training procedures for the DPMD and Deep Wannier neural networks, along with the computational details of the DPMD simulations and SCAN0 hybrid DFT calculations. Sec. III presents the calculation of the dielectric permittivity using the dipole fluctuation method, where transverse and longitudinal dipole relaxation times are determined under metallic and insulating electric boundary conditions, respectively. In Sec. IV, we discuss the evaluation of dielectric permittivity using the Kirkwood dipole correlation approach. Finally, in Sec. V, we summarize our findings and outline potential directions for future research.


\begin{figure*}[t]
    \centering
    \includegraphics[width=6.4in]{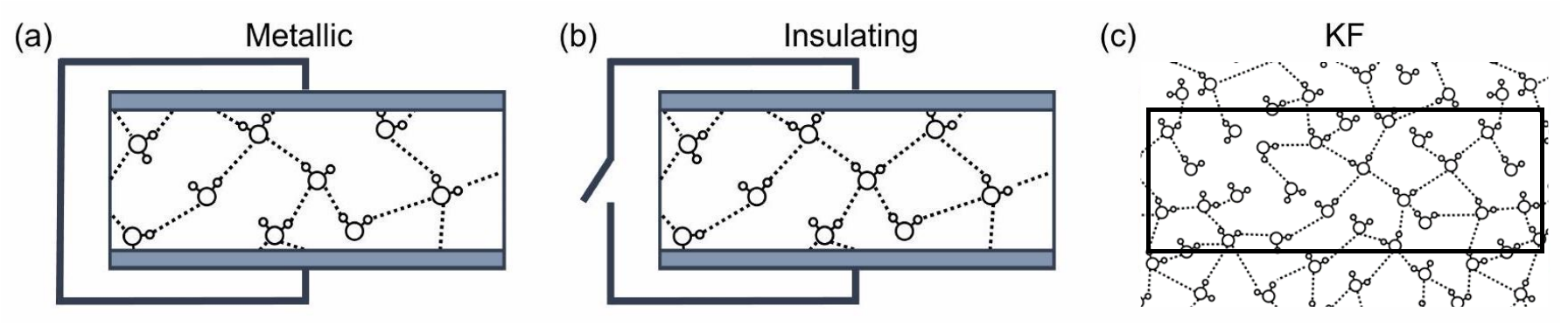}
    \caption{\label{fig:elecboundcond} Schematic diagram of (a) metallic, (b) insulating, and (c) Kirkwood-Fr\"ohlich (KF) electrical boundary conditions.
    }
\end{figure*}

\section{Computational details}

\subsection{DFT Calculations}

DFT calculations were performed using the strongly constrained and appropriately normed (SCAN)~\cite{sun2015strongly} and hybrid SCAN0 exchange-correlation functional as implemented in the Quantum ESPRESSO (QE) package~\cite{giannozzi2017advanced}. 
Electron–nuclei interactions were treated using Hamann--Schl\"{u}ter--Chiang--Vanderbilt (HSCV) pseudopotentials~\cite{hamann1979norm, vanderbilt1985optimally}, with a plane-wave cutoff energy of 150 Ry.
Following self-consistent field convergence, maximally localized Wannier functions (MLWFs)~\cite{marzari1997maximally} were computed using the Wannier90 code~\cite{pizzi2020wannier90}. The 1s electron of hydrogen (H) and $\mathrm{2s^22p^4}$ electrons of oxygen (O) were treated explicitly as valence electrons. Each MLWF was assigned to its nearest oxygen atom, resulting in four doubly occupied MLWFs per oxygen.
The coordinates of the Wannier centroids (WCs) relative to their corresponding atoms were obtained by averaging the positions of the Wannier centers associated with each oxygen atom~\cite{zhang2022deep}.

\subsection{Deep Neural Network (DNN) Models}

Within the DPLR framework, the training dataset was constructed using an active machine learning strategy implemented through the Deep Potential Generator (DP-GEN)~\cite{zhang2019active}. Two DNN models were trained sequentially: the Deep Wannier (DW) model and the DPLR model.
The DW model is designed to predict the coordinates of WCs relative to their corresponding atoms, while the DPLR model learns the potential energy surface with an explicit treatment of long-range electrostatic interactions via Coulombic terms. The procedures for dataset generation and model training are detailed below.

An initial dataset comprising $\sim 1000$ configurations was extracted from a SCAN-based path-integral ab initio molecular dynamics (PI-AIMD) trajectory of 64 water molecules, as reported in Ref.\cite{zhang2021modeling}. For each configuration, the total potential energy $E$, atomic forces $\boldsymbol{F}_i$, and the coordinates of the WCs relative to their corresponding oxygen atoms were computed using the QE package\cite{giannozzi2017advanced} and the Wannier90 code~\cite{pizzi2020wannier90}, employing the DFT settings described previously. These quantities, along with the atomic positions, formed the initial training dataset. 
Based on this dataset, four independent DP models~\cite{zhang2018deep} were trained using the DeePMD-kit~\cite{wang2018deepmd}, each initialized with different random parameters. Model parameters were optimized using the Adam algorithm~\cite{kingma2014adam} with a composite loss function 
\begin{equation}
    \mathcal{L} \left( p_{\mathcal{E}}, p_f \right) = p_{\mathcal{E}} \Delta\mathcal{E}^2 + \frac{p_f}{3n} \sum_i \left|\Delta \boldsymbol{F}_i\right| ,
\end{equation}
where $\Delta \mathcal{E}$ and $\Delta \boldsymbol{F}_i$ denote the discrepancies between the reference data and the current DP predictions for the per-atom energy $\mathcal{E} = E/n$ and atomic forces $\boldsymbol{F}_i$, respectively, where $n$ is the number of atoms. The coefficients $p_{\mathcal{E}}$ and $p_f$ are tunable weighting factors in the loss function. During training, $p_{\mathcal{E}}$ gradually increases from 0.02 to 1, while $p_f$ simultaneously decreases from 1000 to 1, enabling a balanced optimization of energy and force accuracy.

To explore the potential energy surface, PI-DPMD simulations were performed in the isobaric-isothermal ($NPT$) ensemble at 1 bar and 330K or 300 K with 64 water molecules using the i-PI code~\cite{ceriotti2014pi} interfaced with the DeePMD-kit package~\cite{wang2018deepmd}.
At each time step, all four independently trained DP models were used to evaluate the atomic forces, while only one randomly selected model was employed to propagate the trajectory. As the simulations progress, differences among the four models in their predictions of energy $E$ and forces $\boldsymbol{F}_i$ inevitably emerge. To assess model consistency and convergence, we monitored the maximum standard deviation of the predicted forces, defined as $\zeta = \max_i \sqrt{ \langle \left|\boldsymbol{F}_i - \bar{\boldsymbol{F}}_i\right|^2 \rangle }$, where $\bar{\boldsymbol{F}_i} = \langle \boldsymbol{F}_i \rangle$ denotes the average force on atom $i$ over the four models.
The DP models were considered converged when the fraction of configurations with $\zeta > 0.2$ eV/\AA{} dropped below $0.005\%$ of the total PI-DPMD configurations. If this threshold was not met, 50-400 high-discrepancy configurations ($\zeta > 0.2$ eV/\AA{}) were selected, and their energies, atomic forces, and WC coordinates were computed using the aforementioned DFT protocol. These new data were added to the training set, and the model training cycle was repeated until convergence was achieved.

Following dataset construction, the DW model was trained using the atomic configurations and their corresponding WC coordinates as input. The DW model learns to predict the positions of WCs relative to their associated atoms, effectively capturing the local electronic environment. Subsequently, the DPLR model was trained using the same set of atomic configurations, along with their corresponding total energies and atomic forces. The DPLR model explicitly incorporates long-range electrostatic interactions by utilizing the WC coordinates predicted by the pretrained DW model. This hierarchical training strategy enables the DPLR model to accurately reproduce the potential energy surface with an explicit and physically grounded treatment of electrostatics.

\subsection{Classical DPLR-MD Simulations}

The SCAN0-based DPLR model was applied to conduct DPLR-MD simulations in the $NPT$ ensemble at 1 bar and 330 K with a simulation cell containing 4096 water molecules unless otherwise specified. The DPLR-MD simulations were performed by using LAMMPS~\cite{plimpton1995fast} interfaced with the DeePMD-kit package~\cite{wang2018deepmd} for a total duration of 20 ns, with the initial 100 ps excluded from analysis to allow for thermal and structural equilibration.
The DW model was subsequently used to compute the dielectric properties of the simulated systems. Specifically, it was applied to infer the instantaneous dipole moments of individual water molecules from the atomic configurations generated during the DPLR-MD simulations. These molecular dipoles were then used to evaluate macroscopic polarization and its fluctuations, which form the basis for calculating the static dielectric constant and related electrostatic response functions. This approach enables an accurate and efficient characterization of dielectric behavior, leveraging the electronic structure information encoded in the DW model.

\section{Dipole Fluctuation Method}
\begin{equation}
    \frac{4\pi \langle\mathbf{M}^2\rangle}{3 V k_{\mathrm{B}} T} = 
    \frac{\left(\varepsilon - 1\right)\left(2 \varepsilon^{\prime} + 1\right)}{2 \varepsilon^{\prime} + \varepsilon} = 
    \begin{cases}
        \varepsilon - 1, & \left(\varepsilon^{\prime} = \infty\right) \\ 
        1 - \frac{1}{\varepsilon}, & \left(\varepsilon^{\prime} = 0\right) \\ \frac{\left(\varepsilon - 1\right)\left(2 \varepsilon + 1\right)}{3 \varepsilon}, & \left(\varepsilon^{\prime} = \varepsilon\right)
    \end{cases} \label{eq:dipolefluctuation}
\end{equation}

For the liquid water modeled by molecular dynamics in periodic simulation cells containing $N$ water molecules, the static dielectric constant $\varepsilon$ of water can be calculated via the fluctuation of the overall electric dipole moment $\mathbf{M}$ as presented in Eq.~\ref{eq:dipolefluctuation} based on linear response theory. In Eq.~\ref{eq:dipolefluctuation}, $V$, $k_{\mathrm{B}}$, and $T$ are the volume of the simulation cell, Boltzmann constant, and temperature, respectively. In analogy to the experimental measurements, a theoretical simulation of polar liquid demands the specification of its electric boundary condition that being applied on water. In liquid water, the Hamiltonian, thermodynamic enthalpy energy, and predicted integral of motion are all crucially dependent on its electric boundary conditions. According to the Onsager reaction field theory~\cite{ballenegger2004structure}, the dependence of the dielectric constant of an isotropic liquid on its electric boundary condition can be simply determined by one parameter $\varepsilon^{\prime}$ in Eq.~\ref{eq:dipolefluctuation}.

In the above, the electric boundary is described by a macroscopic sample of liquid water of dielectric constant $\varepsilon$ surrounded by a homogeneous medium with a dielectric constant of $\varepsilon^{\prime}$. Typically, there are two electric boundary conditions adopted in molecular dynamics simulations, namely insulating ($\varepsilon^{\prime} = 0$) and metallic ($\varepsilon^{\prime} = \infty$) boundary conditions, which correspond to the open-circuit and closed-circuit settings in experimental measurements in Fig.~\ref{fig:elecboundcond} (a, b).

In addition, a third electric boundary condition is often applied for theoretical discussions under which the macroscopic sample of water is embedded in a water reservoir ($\varepsilon^{\prime} = \varepsilon$) as schematically shown in Fig.~\ref{fig:elecboundcond} (c). It was first introduced by Kirkwood-Fr\"ohlich (KF)~\cite{kirkwood1939dielectric, frohlich1958theory} to extract the correlation length among collective dipole fluctuations on a H-bond network (see discussions in Sec. IV). However, in practice, the implementation of KF boundary condition requires the knowledge of water's dielectric constant beforehand and therefore cannot be applied to predict the dielectric constant of water. Under the metallic, insulating, and KF electric boundary conditions, the formula of computing the $\varepsilon$ in water are further simplified as shown on the right hand side of Eq.~\ref{eq:dipolefluctuation}.

\subsection{Dipole fluctuation under metallic boundary condition}
Within the framework of the dipole fluctuation approach, the metallic electric boundary condition has been the most applied one to evaluate the dielectric constant in liquid water because most classical MD and AIMD simulations adopt Ewald summation technique in computing the electrostatic potential energies. In Ewald summation, the applied electric field on a system is constrained to zero by the vanishing electric potential drop across the macroscopic sample, which is equivalent to applying the metallic boundary condition. Under the metallic boundary condition, the dipole fluctuation of water is transverse in nature. Therefore, a converged simulation of transverse dipole fluctuation in water is essential for an accurate prediction of its static dielectric constant. As shown in Fig.~\ref{fig:dipolefluctepsilon} (a),
we present the electric dipole $\mathbf{M}$ as a function of simulation time along one of the orthogonal directions in the DPMD simulations. Clearly, it can be seen in Fig.~\ref{fig:dipolefluctepsilon} (a) that $\mathbf{M}$ has a mean value $\mathbf{M} = 0$, which is expected since water is a paraelectric material with vanishing net polarization at room temperature. Nevertheless, $\mathbf{M}$ undergoes a large fluctuation and a significant second moment $\left\langle\mathbf{M}^2\right\rangle$ giving rise to its large static dielectric constant. However, $\left\langle\mathbf{M}^2\right\rangle$ converges rather slowly with simulation time as shown in Fig.~\ref{fig:dipolefluctepsilon} (c), which is due to the slow relaxation of polarization. According to Debye's theory of dielectric relaxation, the polarization fluctuation decays exponentially over time whose characteristic time $\tau^{\epsilon'=\infty}_{\mathbf{M}}$ can be determined from the autocorrelation $C_{\mathbf{M}}\left(t\right)$ of the electric dipole in the system defined as $C_{\mathbf{M}}\left(t\right)=\frac{\left\langle \mathbf{M}\left(t\right) \cdot \mathbf{M}\left(0\right) \right\rangle}{\left\langle\mathbf{M}\left(0\right)^2\right\rangle}$. In our simulation, a relaxation time $\tau^{\epsilon'=\infty}_{\mathbf{M}}=8.1$ $ps$ has been extracted from our DPMD simulation, which is close to the experimentally reported Debye relaxation time of 8.28 $ps$ at room temperature~\cite{ellison2007permittivity}. In the MD computer simulations, the simulation time $t$ should be long enough to sample sufficient cycles of polarization relaxation in water for a well-converged dipole fluctuation $\left\langle\mathbf{M}^2\right\rangle$. The formulas of the relative error $I_{\varepsilon - 1}$ in the predicted dielectric constant are shown in Eq.~\ref{eq:relativeerror} in which the accuracy is inversely proportional to $\frac{t}{\tau_{\mathbf{M}}}$ as expected. Because of the slow Debye relaxation, it has been widely noticed that simulation time at the scale of tens of nanoseconds is required to have a converged prediction of the static dielectric constant of water. In our DPMD simulation over 20 $ns$ , the dielectric constant of water is determined to be $\varepsilon = 102$ with a small relative error of $3 \%$ according to Eq.~\ref{eq:relativeerror}~\cite{ballenegger2004structure}.

\begin{equation}
    I_{\varepsilon - 1} = 
    \frac{2 \varepsilon^{\prime} + \varepsilon}{2 \varepsilon^{\prime} + 1} \sqrt{\frac{2 \tau^{\epsilon'}_{M}}{3 t}} = 
    \begin{cases}
        \sqrt{\frac{2 \tau^{\epsilon'=\infty}_{M}}{3 t}}, & \left(\varepsilon^{\prime} = \infty\right) \\ 
        \varepsilon \sqrt{\frac{2 \tau^{\epsilon'=0}_{M}}{3 t}}, & \left(\varepsilon^{\prime} = 0\right) \\ 
        \frac{3 \varepsilon}{2 \varepsilon + 1} \sqrt{\frac{2 \tau^{\epsilon'=\epsilon}_{M}}{3 t}}, & \left(\varepsilon^{\prime} = \varepsilon\right)
    \end{cases}     \label{eq:relativeerror}
\end{equation}

\subsection{Dipole fluctuation under insulating boundary condition}
\begin{figure*}[!t]
  \centering
    \includegraphics[width=5.5 in]{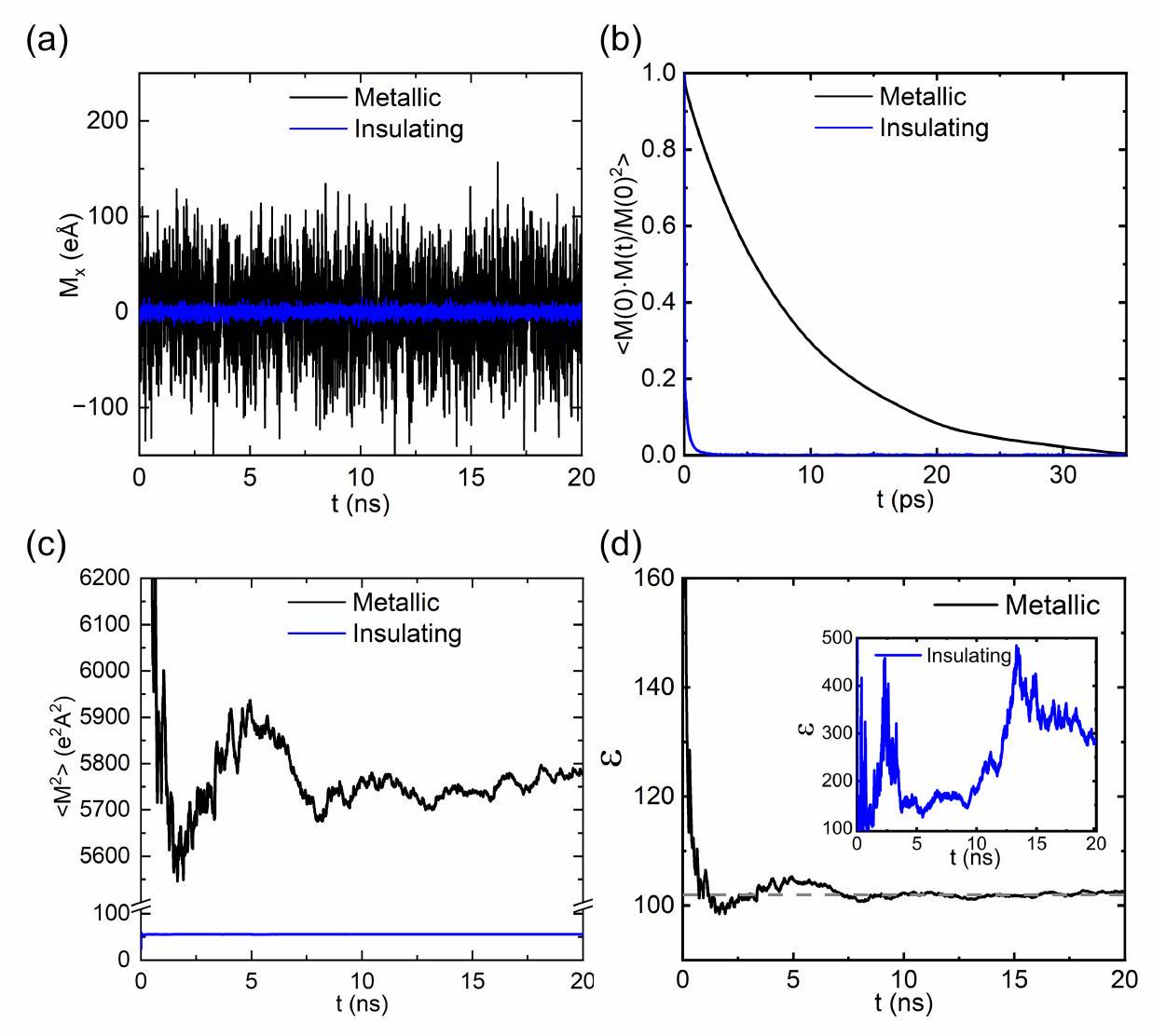}
    \caption{\label{fig:dipolefluctepsilon} (a) Time evolution of the total dipole moment $M_x$ in the $x$ direction under different electrical boundary conditions. 
    (b) Time autocorrelation function of the total dipole moment under different electrical boundary conditions. 
    (c) Time accumulative average of dipole fluctuation under different electrical boundary conditions. (d) Time accumulative average of dielectric constants calculated using dipole fluctuation formula under different electrical boundary conditions.
    }
\end{figure*}

The insulating electric boundary condition is equivalent to an open-circuit capacitor setup in experiments as schematically shown in Fig.~\ref{fig:elecboundcond} (b), in which no free charge exists on the metal plates. Because of the vanishing free charge on the capacitor, a constrained zero electric displacement field is imposed on the macroscopic sample of water~\cite{stengel2009electric}. Under the insulating boundary condition, the dipole fluctuation is longitudinal in nature. In comparison with the transverse dipole fluctuation under metallic boundary condition, the longitudinal dipole fluctuation is largely suppressed as shown in Fig.~\ref{fig:dipolefluctepsilon} (a). The mechanism is similar to that behind the experimentally observed higher phonon energies of longitudinal optical phonons compared to the transverse optical in ionic solid materials~\cite{ashcroft1976solid}. In addition to the suppressed dipole fluctuation, the polarization relaxation time $\tau^{\epsilon'=0}_{\mathbf{M}}$ is greatly reduced as well. In a large band gap insulator such as neat water, no excess free charge $\rho_{\text{free}}$ exists, therefore $\nabla \cdot \mathbf{D}=\rho_{\text{free}}=0$ resulting in a uniform electric displacement field $\mathbf{D}$ throughout the system. The application of insulating electric boundary condition in DPMD simulation enables us to treat the long-range Coulombic interaction by the uniform parameter $\mathbf{D}$ throughout the neat water. Under this treatment, the dipole fluctuation in water becomes nearsighted and less correlated in real space, which results in a largely suppressed polarization relaxation time $\tau^{\epsilon'=0}_{\mathbf{M}}$. As shown in Fig.~\ref{fig:dipolefluctepsilon} (b), the polarization relaxation time $\tau^{\epsilon'=0}_{\mathbf{M}}=0.3$ ps is obtained in our computer simulations under the insulating electric boundary condition, and this value is in close agreement with the recent result reported by Sprik \emph{et al.}~\cite{zhang2016computing}. Under insulating electric boundary condition, the fast polarization relaxation time $\tau^{\epsilon'=0}_{\mathbf{M}}$ also helps the quicker convergence of $\left\langle\mathbf{M}^2\right\rangle$ as compared to that under metallic boundary conduction as seen in Fig.~\ref{fig:dipolefluctepsilon} (c). Because of the faster polarization relaxation time $\tau_{\mathbf{M}}^{\varepsilon'}$ in insulating boundary condition, one expects that the predicted static dielectric constant of water should converge much faster than that in metallic boundary condition. However, our DPMD simulation suggests the opposite. As can be clearly seen in Fig.~\ref{fig:dipolefluctepsilon} (d), the predicted dielectric at the time scale of 20 $ns$ is barely converged with significant error bars. This seemingly surprising result is due to the fact that the prediction of $\varepsilon$ under insulating boundary condition demands the inversion of a small numerical number in Eq.~\ref{eq:dipolefluctuation} as $\varepsilon = \frac{1}{1 - \frac{4 \pi 
\left\langle\mathbf{M}^2\right\rangle}{3 V k_{\mathrm{B}} T}}$ which poses an additional challenge on the accuracy of predicted $\varepsilon$. The above effect can be more easily seen in the analytic formula of the relative errors in Eq. 2. The relative error of the insulating boundary condition has an extra term of $\varepsilon$ reflecting the loss of accuracy by inverting a small number. Based on Eq.~\ref{eq:relativeerror}, it can be estimated that a DMPD trajectory over 2.2 microseconds ($\mu s$) is required in order to achieve a similar accuracy as reported for a DPMD simulation performed under the metallic boundary condition within 20 ns.
\section{Dipole Correlation Method}
\begin{figure*}[t]
    \centering
    \includegraphics[width=5.6in]{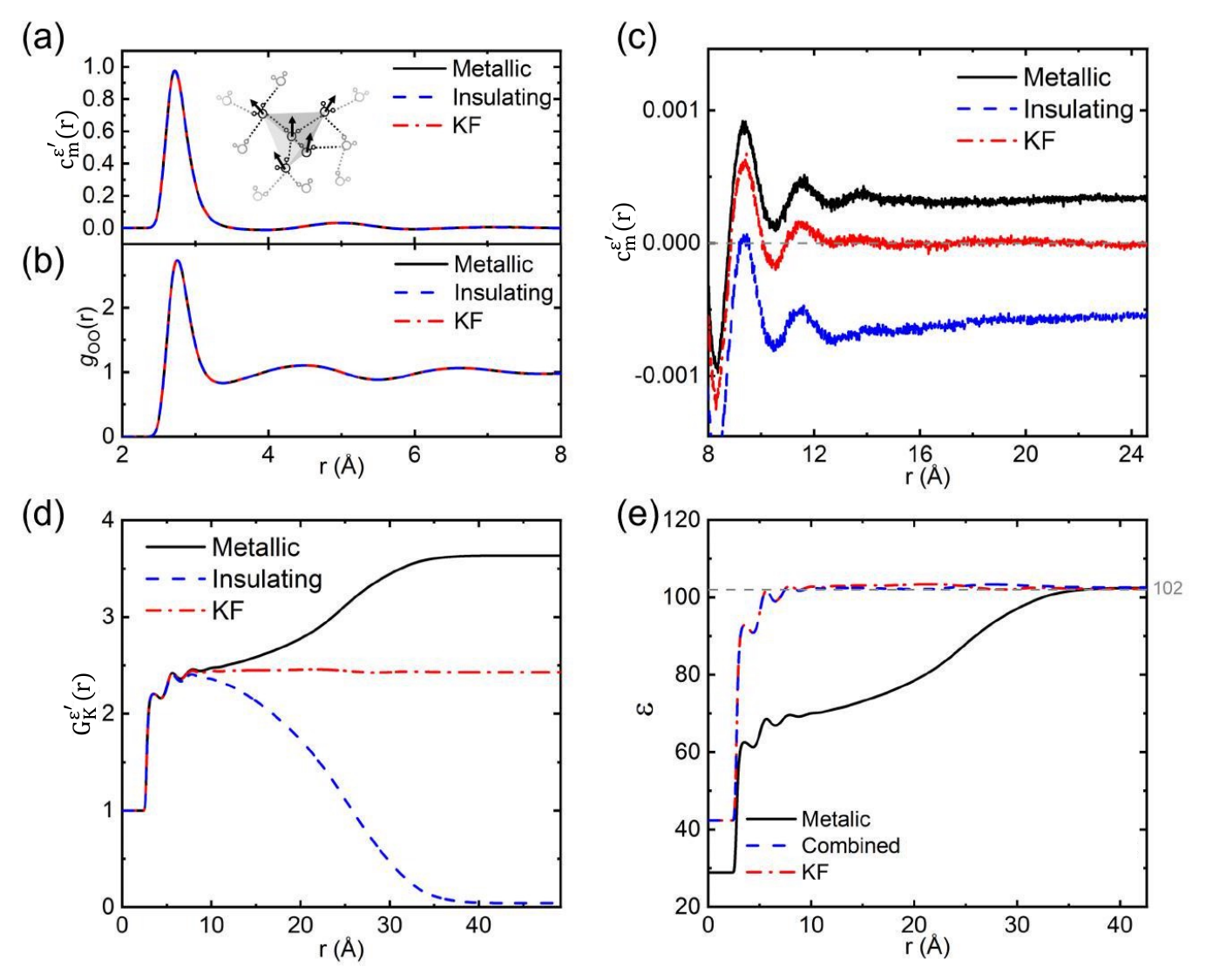}
    \caption{\label{fig:dipolecorrelation} (a) Dipolar correlation function, $c_{m}^{\varepsilon'}\left(r\right)$, and 
    (b) Oxygen-oxygen radial distribution function, $g_{\mathrm{OO}}\left(r\right)$, and simulated under different electrical boundary conditions with a simulation cell containing 4096 water molecules. The inset in (a) schematically illustrates the tetrahedral hydrogen bond network in liquid water. 
    (c) Enlarged view of the dipolar correlation function, $c_{m}^{\varepsilon'}\left(r\right)$, showing the long-range behavior of $c_{m}^{\varepsilon'}\left(r\right)$. 
    (d) Correlation factor $G_{\mathrm{K}}^{\varepsilon'}\left(r\right)$ calculated by integrating the dipolar correlation function, $c_{m}^{\varepsilon'}\left(r\right)$, under different electrical boundary conditions.
    (e) Dielectric constant $\varepsilon$ calculated using the correlation formula under different electrical boundary conditions.
    }
\end{figure*}

Alternatively, the dipole correlation method is also an important approach often applied to calculate the static dielectric constant in water. In the seminal paper in 1939, Kirkwood pointed out that the dipole fluctuation in water is collective in nature~\cite{kirkwood1939dielectric}. The collective dipole fluctuation is facilitated by the angular correlation among the water molecules, and for a pair of water molecules on water's H-bond network, its dipolar correlation rapidly decays to zero as the distance between them increases in real space. Based on the above, Kirkwood~\cite{kirkwood1939dielectric} further proposed that the dielectric constant in liquid water can be simply computed with the knowledge of a single number, Kirkwood correlation factor $G_{\mathrm{K}}^{\varepsilon'}$, which accounts for the correlated dipole fluctuations in water. When the correlation method was proposed, Kirkwood modeled a macroscopic water droplet embedded inside a water reservoir whose boundary condition is referred to as the KF electric boundary condition. Later, Fr\"ohlich adapted Onsager's reaction and provided the formula under arbitrary electric boundary conditions as shown in Eq.~\ref{eq:KFeqn}, which is known as the Kirkwood-Fr\"ohlich equation.
\begin{equation}
    \frac{4 \pi \rho \mu^{2} G^{\epsilon'}_{\mathrm{K}}}{3 k_{\mathrm{B}} T} = 
    \frac{\left(\varepsilon - 1\right)\left(2 \varepsilon^{\prime} + 1\right)}{2 \varepsilon^{\prime} + \varepsilon} = 
    \begin{cases}
        \varepsilon - 1, & \left(\varepsilon^{\prime} = \infty\right) \\
        1 - \frac{1}{\varepsilon}, & \left(\varepsilon^{\prime} = 0\right) \\ \frac{\left(\varepsilon - 1\right)\left(2 \varepsilon + 1\right)}{3 \varepsilon}, & \left(\varepsilon^{\prime} = \varepsilon\right)
    \end{cases}     \label{eq:KFeqn}
\end{equation}
Instead of directly evaluating the dipole fluctuation, the Kirkwood-Fr\"ohlich equation in Eq.~\ref{eq:KFeqn} computes the dielectric constant via two constants under a statistical ensemble, which are the correlation factor $G^{\epsilon'}_{\mathrm{K}}$ and the average dipole moment $\mu$ per water molecule. In Eq.~\ref{eq:KFeqn}, the correlation factor $G^{\epsilon'}_{\mathrm{K}}$ is a constant in liquid water under an applied electric boundary condition controlled by the variable $\epsilon'$. The numerical value of $G^{\epsilon'}_{\mathrm{K}}$ is determined by the volume integral of the dipolar correlation function $c_{m}^{\varepsilon'}\left(r\right)$ over the entire simulation box as presented in Eq.~\ref{eq:GK}.
\begin{align}
    G_{\mathrm{K}}^{\varepsilon'} &= 1 + \rho \int c_{m}^{\varepsilon'}\left(r\right) d \boldsymbol{r} ,    \label{eq:GK} \\
    c_{m}^{\varepsilon'}\left(r\right) &= \frac{1}{\rho N} \left\langle \sum_{i \neq j} \hat{\boldsymbol{\mu}}_{i} \cdot \hat{\boldsymbol{\mu}}_{j} \delta \left( \boldsymbol{r} + \boldsymbol{r}_{i} - \boldsymbol{r}_{j} \right) \right\rangle .   \label{eq:cmr}
\end{align}
In the above, the dipolar pair correlation function $c_{m}^{\varepsilon'}\left(r\right)$, as defined in Eq.~\ref{eq:cmr}, is an ensemble function of distance $r$ normalized by the molecule number density $\rho$ and total number of molecules $N$ in the simulation cell, which accounts for the angular correlation between the unit vector of two dipole moments $\hat{\boldsymbol{\mu}}_{i}$ and $\hat{\boldsymbol{\mu}}_{j}$ on two distinct water molecules located at distances of $\boldsymbol{r}_{i}$ and $\boldsymbol{r}_{j}$ from the reference molecule at the origin point, respectively.

It should be noted that the dipole fluctuation and dipole correlation approaches are equivalent to each other. Indeed, the correlation formalism of Eq.~\ref{eq:KFeqn} can be derived straightforwardly from the dipole fluctuation formalism of Eq.~\ref{eq:dipolefluctuation} by replacing the total electric dipole $\mathbf{M}$ as the vector sum of individual dipoles $\boldsymbol{\mu}_{i}$ of water molecules $\mathbf{M}=\sum_{i=1}^{N} \boldsymbol{\mu}_{i}$ and using the property of neat water as a homogeneous polar liquid $\left\langle\boldsymbol{\mu}_{i} \cdot \boldsymbol{\mu}_{i}\right\rangle = \mu^2$. Nevertheless, the correlation approach has a clear advantage that it allows us to carry out analysis in real space on the dipolar correlation function and to understand the nature of correlated dipole fluctuation in the calculated static dielectric constant $\varepsilon$. As shown in Fig.~\ref{fig:dipolecorrelation} (a),
we present the simulated dipolar pair correlation function $c_{m}^{\varepsilon'}\left(r\right)$ under metallic, insulating, and KF electric boundary conditions. For clarity, the $c_{m}^{\varepsilon'}\left(r\right)$ functions are presented at short-range and long-range scales of distance $r$ in Fig.~\ref{fig:dipolecorrelation} (a) and Fig.~\ref{fig:dipolecorrelation} (c), respectively.
\subsection{Short-Range Dipole Correlations in the H-Bond Network}
At short range, it can be clearly seen in Fig.~\ref{fig:dipolecorrelation} (a) that two water dipoles are strongly correlated at short-range scales in the H-bond network. The dipolar pair correlation functions $c_{m}^{\varepsilon'}\left(r\right)$ under all three different electric boundary conditions are in quantitative agreement with each other in Fig.~\ref{fig:dipolecorrelation} (a), which are featured by pronounced positive peaks in the functions of $c_{m}^{\varepsilon'}\left(r\right)$. For comparison, the oxygen-oxygen pair distribution function $g_{\mathrm{OO}}\left(r\right)$ is also presented in Fig.~\ref{fig:dipolecorrelation} (b). The close similarity between $c_{m}^{\varepsilon'}\left(r\right)$ in Fig.~\ref{fig:dipolecorrelation} (a) and $g_{\mathrm{OO}}\left(r\right)$ in Fig.~\ref{fig:dipolecorrelation} (b) indicates that the features in $c_{m}^{\varepsilon'}\left(r\right)$ are also determined by the structure of the H-bond network in water, the same mechanism behind the coordination structures observed in $g_{\mathrm{OO}}\left(r\right)$. Due to the $sp^3$ hybridization as its electronic configuration, water molecules construct a near-tetrahedral H-bond network. A water molecule at the center of a tetrahedron is likely to donate two H-bonds to and accept two H-bonds from neighboring water molecules at the four vertices of the tetrahedron as depicted in the inset of Fig.~\ref{fig:dipolecorrelation}(a). Because the H-bonding is directional in nature, the electric dipoles on the H-bonded water molecules in the first coordination shell are also aligned roughly in the same direction as the dipole on the water molecule at the origin, which gives rise to the first sharp peak observed in $c_{m}^{\varepsilon'}\left(r\right)$. The water molecules in the second coordination shell are not directly H bonded to the water molecule at the origin, but their dipole orientation is still largely influenced and aligned by the dipole at the central molecule via the near-tetrahedral H-bond network, which yields the second positive peak at $r=5$ \AA{} in the function of $c_{m}^{\varepsilon'}\left(r\right)$. As distance increases, the two dipoles between molecules at $r$ and the water molecule at the origin become less and less correlated due to the disordered liquid structure under thermal fluctuation. Consistently, the positive correlation peak decays in the third coordination shell and beyond as shown in Fig.~\ref{fig:dipolecorrelation} (a, c). On the other hand, the region between the first and second coordination shells is populated with interstitial water molecules that are not H-bonded. These dipoles on the non-bonded interstitial water molecules are slightly anti-parallel to the dipole on the water molecule at the origin, resulting in a negative correlation in this region as shown in Fig.~\ref{fig:dipolecorrelation} (a).
\subsection{Long-Range Dipole Correlations in the H-Bond Network}
As the distance $r$ between two water molecules further increases, the oscillation features in both the oxygen-oxygen pair distribution function $g_{\mathrm{OO}}\left(r\right)$ of Fig.~\ref{fig:dipolecorrelation} (b) and the dipolar correlation function $c_{m}^{\varepsilon'}\left(r\right)$ in Fig.~\ref{fig:dipolecorrelation} (a) fade out, which indicates the disappearance of coordination shells and weaker dipolar correlations between pairs of water molecules in the H-bond network. However, the function of $c_{m}^{\varepsilon'}\left(r \rightarrow \infty\right)$ at the long-range limit does not vanish completely; instead, it decays to small constants dependent on its electric boundary conditions as shown in Eq.~\ref{eq:longrangecmr}.

\begin{equation}
    \begin{aligned}
        \lim_{r \rightarrow \infty} c_{m}^{\epsilon'}\left(r\right) 
        & = \frac{1}{2 \pi \beta \rho^{2} \mu^{2} V} \frac{\left(\varepsilon - 1\right)^2 \left(\varepsilon^{\prime} - \varepsilon\right)}{\varepsilon\left(2 \varepsilon^{\prime} + \varepsilon\right)} \\
        & = \begin{cases}
            \frac{1}{4 \pi \beta \rho^2 \mu^2 V} \frac{\left(\varepsilon - 1\right)^2}{\varepsilon}, & \left(\varepsilon^{\prime} = \infty\right) \\
            -\frac{1}{2 \pi \beta \rho^2 \mu^2 V} \frac{\left(\varepsilon - 1\right)^{2}}{\varepsilon}, & \left(\varepsilon^{\prime} = 0\right) \\
            0, & \left(\varepsilon^{\prime} = \varepsilon\right)
        \end{cases}
    \end{aligned}   \label{eq:longrangecmr}
\end{equation}

As shown in Fig.~\ref{fig:dipolecorrelation} (c), distinct features of $c_{m}^{\varepsilon'}\left(r \rightarrow \infty\right)$ under metallic, insulating, and KF boundary conditions can be identified. In comparison with the $c_{m}^{\varepsilon'}\left(r\right)$ at short range as determined by the H-bonding, the above long-range behavior of $c_{m}^{\varepsilon'}\left(r\right)$ is instead governed by classical electromagnetism.

\begin{figure}[t]
    \centering
    \includegraphics[width=3.3in]{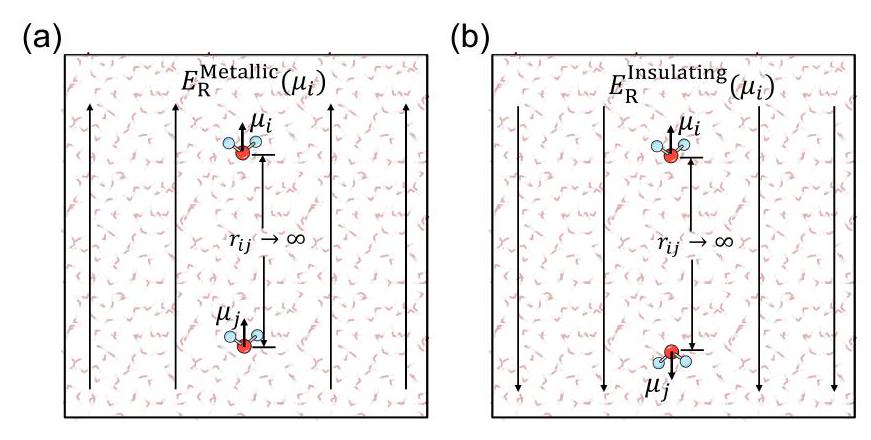}
    \caption{\label{fig:reactionfieldelecboundcond} Schematic diagram of the reaction field produced by $\boldsymbol{\mu}_{i}$ under (a) metallic and (b) insulating electrical boundary conditions.
    }
\end{figure}

At the macroscopic scale, a water molecule in the simulation box can be considered as a point dipole embedded in a homogeneous dielectric medium, and its electrostatic interaction with surroundings is facilitated by the Coulomb field of the screened point dipole $\boldsymbol{\mu}_{i}^{\text{scr}}=\frac{3 k_{\mathrm{B}} T \left(\varepsilon - 1\right)}{4 \pi \rho \mu^2 \varepsilon} \boldsymbol{\mu}_{i}$. As the dipole field propagates in the dielectric medium, surface charges will emerge at the boundary due to the dielectric discontinuity $\varepsilon \neq \varepsilon^{\prime}$ that occurs there. Based on the Onsager reaction field theory, the overall effect of the surface charges can be conveniently represented by a constant applied electric field $\boldsymbol{E}_{R,i}^{\varepsilon'}$ inside the system as shown in Eq.~\ref{eq:reactionfield}.
\begin{equation}
    \boldsymbol{E}_{R,i}^{\varepsilon'} = 
    \frac{2 \left(\varepsilon^{\prime} - \varepsilon\right) \boldsymbol{\mu}_{i}^{\mathrm{scr}}}{\varepsilon \left(2 \varepsilon^{\prime} + \varepsilon\right) V} = 
    \begin{cases}
        \frac{\mu_{i}^{\mathrm{scr}}}{\varepsilon V}, & \left(\varepsilon^{\prime} = \infty\right) \\
        -\frac{2 \mu_{i}^{\mathrm{scr}}}{\varepsilon V}, & \left(\varepsilon^{\prime} = 0\right) \\ 
        0, & \left(\varepsilon^{\prime} = \varepsilon\right)
    \end{cases}     \label{eq:reactionfield}
\end{equation}
In the above, $\boldsymbol{E}_{R,i}^{\varepsilon'}$ is the well-known reaction field, and its expression can be easily derived by solving Laplace's equation in a continuous medium. The magnitude of the reaction field $\boldsymbol{E}_{R,i}^{\varepsilon'}$ is inversely proportional to the volume $V$ of the simulation cell. However, the direction of $\boldsymbol{E}_{R,i}^{\varepsilon'}$ crucially depends on the electric boundary condition as shown in Fig.~\ref{fig:reactionfieldelecboundcond},
in which the direction of $\boldsymbol{E}_{R,i}^{\varepsilon'}$ is parallel (anti-parallel) to the direction of the screened point dipole $\boldsymbol{\mu}_{i}^{\text{scr}}$ under metallic (insulating) boundary conditions. The presence of $\boldsymbol{E}_{R,i}^{\varepsilon'}$ determines the asymptotic behavior of $c_{m}^{\varepsilon'}\left(r \rightarrow \infty\right)$. Now we consider two water molecules embedded in the water reservoir, as schematically shown in Fig.~\ref{fig:reactionfieldelecboundcond}, which are separated by macroscopic distance $r$ in liquid water. This pair of water molecules will interact with each other via the screened point dipoles of $\boldsymbol{\mu}_{i}^{\text{scr}}$ and $\boldsymbol{\mu}_{j}^{\text{scr}}$, respectively. In this scenario, the $\boldsymbol{\mu}_{i}^{\text{scr}}$ generates a reaction field $\boldsymbol{E}_{R,i}^{\varepsilon'}$ in the system, which will act on the $\boldsymbol{\mu}_{j}^{\text{scr}}$ via the coupling energy term of $\boldsymbol{E}_{R,i}^{\varepsilon'} \cdot \boldsymbol{\mu}_{j}^{\text{scr}}$. Under metallic boundary conditions, the coupling $\boldsymbol{E}_{R,i}^{\varepsilon'} \cdot \boldsymbol{\mu}_{j}^{\text{scr}}$ scales as $\left(\boldsymbol{\mu}_{i}^{\text{scr}} \cdot \boldsymbol{\mu}_{j}^{\text{scr}}\right) / V$, which results in a positive constant correlation $c_{m}^{\varepsilon'}\left(r \rightarrow \infty\right)$ as shown in Fig.~\ref{fig:dipolecorrelation} (c). On the contrary, a negative constant dipolar correlation function $c_{m}^{\varepsilon'}\left(r \rightarrow \infty\right)$ is observed in Fig.~\ref{fig:dipolecorrelation} (c) under insulating boundary conditions because the coupling between the reaction field and the screened dipole $\boldsymbol{\mu}_{j}^{\text{scr}}$ scales as $-\left(\boldsymbol{\mu}_{i}^{\text{scr}} \cdot \boldsymbol{\mu}_{j}^{\text{scr}}\right) / V$. It should be noted that $c_{m}^{\varepsilon'}\left(r \rightarrow \infty\right) = 0$ only when the KF boundary condition is applied, and the $\varepsilon$ is continuous at the boundary, and such a boundary condition can be only modeled in theoretical simulations with prior knowledge of $\varepsilon$ in water. Indeed, our current calculations under the KF electric boundary condition, based on the dipole correlation method, employ the dielectric constant $\varepsilon=102$ computed in Section III by dipole fluctuation method.
\begin{figure*}[ht]
    \centering
    \includegraphics[width=6.0in]{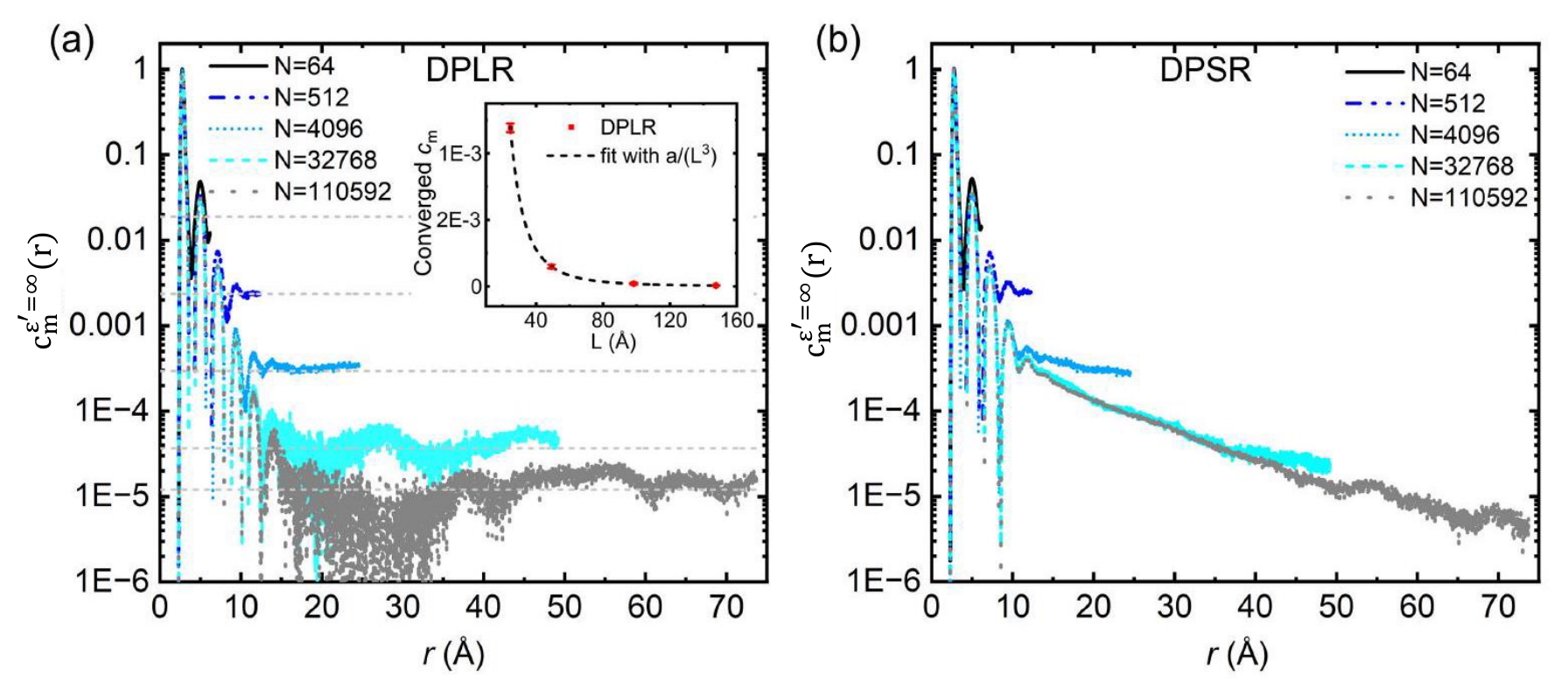}
    \caption{\label{fig:longrangelimitcmr} Log-scale dipolar correlation function, $c_{m}^{\varepsilon'}\left(r\right)$, simulated using (a) DPLR and (b) DPSR, respectively, for different cell sizes with $N$ representing the number of water molecules in the simulation cell. All simulations were performed under the metallic boundary condition. The inset in (a) presents the converged constant of $c_{m}^{\varepsilon'}\left(r\right)$ simulated using DPLR for different cell lengths $L$, which aligns well with the $\sim \frac{1}{L^{3}}$ behavior.
    }
\end{figure*}

The behavior of the dipolar correlation function $c_{m}^{\varepsilon'}\left(r \rightarrow \infty\right)$ is a direct consequence of the long-range Coulomb forces. Therefore, an explicit treatment of the long-range electrostatic forces in the underlying deep neural network is essential to predict the correct asymptotic behavior. As shown in Fig.~\ref{fig:longrangelimitcmr} (a) and (b),
we compare the dipolar correlation functions $c_{m}^{\varepsilon'=\infty}\left(r\right)$ generated by molecular simulations using the long-range Coulomb interaction inclusive DPLR model and the standard short-range DPSR model under the metallic boundary condition. In both DPLR and DPSR based molecular dynamics simulations, the liquid water structure is modeled by various sizes of simulation boxes of lengths $L$ that contain $N$ number of water molecules. Clearly, it can be seen that the DPLR model correctly predicts the constant of $c_{m}^{\varepsilon'=\infty}\left(r\right)$ as $r \rightarrow \infty$. Moreover, the value of $c_{m}^{\varepsilon'=\infty}\left(r \rightarrow \infty\right)$ decays with the volume of the simulation box, which scales as $\sim 1 / V=1 / L^{3}$ as expected, shown in the insert of Fig.~\ref{fig:longrangelimitcmr} (a). In sharp contrast, $c_{m}^{\varepsilon'=\infty}\left(r \rightarrow \infty\right)$ predicted by the DPSR model is qualitatively incorrect, yielding an exponential decay as $r \rightarrow \infty$ for all simulations, as shown in Fig.~\ref{fig:longrangelimitcmr} (b). Indeed, in the absence of long-range Coulombic forces, the dipole correlation function, as an electronic property, exhibits exponential decay with distance from a local perturbation, in accordance with Walter Kohn’s nearsightedness principle~\cite{KohnPNAS}.

\subsection{Kirkwood Correlation Length and Correlation Factor}

Once the asymptotic behavior of the dipolar correlation function $c_{m}^{\varepsilon'}\left(r \rightarrow \infty\right)$ is achieved, the Kirkwood correlation length $r_{c}$ can be computed straightforwardly for each electric boundary condition as aforementioned. As shown in Fig. 3 (c), the Kirkwood correlation length $r_{c}$ can be determined at the distance when two water-dipoles completely lose their angular correlation, i.e., $c_{m}^{\varepsilon'}\left(r > r_{c}\right) = \text{const.}$.
Specifically, we first evaluate the mean and standard deviation of $c_{m}^{\varepsilon'}\left(r\right)$ over a range of long distances where the function exhibits minimal variation. We then scan $c_{m}^{\varepsilon'}\left(r\right)$ from short to long distances and identify the smallest value of $r$ for which $c_{m}^{\varepsilon'}\left(r\right)$ remains within one standard deviation of the asymptotic mean. This criterion provides a robust estimate of the distance scale beyond which long-range dipolar correlations become negligible.
Under each of considered electric boundary conditions, our simulation yields a rather consistent correlation length $r_{c} \sim 16.0$ \AA{} under the KF boundary condition, $r_{c} \sim 16.4$ \AA{} under the metallic boundary condition and $r_{c} \sim 16.7$ \AA{} under the insulating boundary condition. Kirkwood correlation length $r_{c}$ as large as $\sim$ 16 \AA{} suggests that a well-converged prediction of the dielectric constant of water demands the liquid model to contain at least five coordination shells as shown by the function $g_{\rm oo}(r)$ in Fig. 3. In this regard, roughly at least 1100 number of water molecules should be modeled in the model of water under ambient conditions.

With the correct asymptotic behavior of the dipolar correlation function $c_{m}^{\varepsilon'}\left(r \rightarrow \infty\right)$, the Kirkwood correlation factor can be computed via the volume integral of $c_{m}^{\varepsilon'}$, as described in  Eq.~\ref{eq:GK}. However, care must be taken when extracting $G_\mathrm{K}^{\epsilon'}$, as its interpretation depends on the chosen boundary conditions. Strictly speaking, $G_\mathrm{K}^{\epsilon'=\epsilon}$ should be evaluated under the KF boundary condition, as originally formulated, to preserve its clear and intuitive physical meaning: for a system of uncorrelated dipoles, $G_\mathrm{K}^{\epsilon'=\epsilon}=1$. In liquid water, the near-tetrahedral hydrogen-bond network induces local alignment among water dipoles, resulting in $G_\mathrm{K}^{\epsilon'=\epsilon}>1$. The magnitude of $G_\mathrm{K}^{\epsilon'=\epsilon}$ thus serves as a quantitative measure of how strongly correlated dipoles collectively respond to screen an applied electric field in the static limit. In our simulation, we obtain $G_\mathrm{K}^{\epsilon'=\epsilon}=2.43$, which is in close agreement with the value $G_\mathrm{K}^{\epsilon'=\epsilon} \sim 2.5$ estimated by Pople at room temperature~\cite{pople1951molecular}. His estimate was based on the assumption that the primary effect of thermal disorder in liquid water is the bending of hydrogen bonds, while the fourfold coordination characteristic of ice remains largely intact.

Under metallic and insulating electric boundary conditions, the correlation factors $G_{\mathrm{K}}^{\epsilon'=\infty}$ and $G_\mathrm{K}^{\epsilon'=0}$ can also be computed straightforwardly using the same method described in Eq.~\ref{eq:GK}. In the literature, they are sometimes casually referred to as Kirkwood correlation factors as well. However, it is important to emphasize that their values cannot be interpreted in the same way as the true Kirkwood correlation factor $G_\mathrm{K}^{\epsilon'=\epsilon}$, since they are significantly influenced by the choice of electric boundary conditions. Due to the presence of reaction fields, the dipolar correlation functions $c_m^{\epsilon'=\infty}$ and $c_m^{\epsilon'=0}$ do not vanish as $r \rightarrow \infty$ but instead decays to a constant value inversely proportional to the simulation box volume, i.e., $\sim 1/V$, as discussed in the previous subsection. As a result, the correlation factors obtained in our simulations under metallic and insulating boundary conditions are $G_{\mathrm{K}}^{\epsilon'=\infty} = 3.64$ and $G_{\mathrm{K}}^{\epsilon'=0} = 0.04$, respectively—values that differ substantially from the true Kirkwood correlation factor $G_{\mathrm{K}}^{\epsilon'=\epsilon}=2.43$.

{\small
\begin{align}
    G_{\mathrm{K}}^{\mathrm{\epsilon'=\infty}}=G_{\mathrm{K}}^{\mathrm{\epsilon'=\epsilon}} + N c_{m}^{\mathrm{\epsilon'=\infty}} \left(r \rightarrow \infty\right) = G_{\mathrm{K}}^{\mathrm{\epsilon'=\epsilon}} + \frac{1}{4 \pi \beta \rho \mu^2} \frac{\left(\varepsilon - 1\right)^2}{\varepsilon} , \label{eq:GKmetallic} \\
    G_{\mathrm{K}}^{\mathrm{\epsilon'=0}}=G_{\mathrm{K}}^{\mathrm{\epsilon'=\epsilon}} + N c_{m}^{\mathrm{\epsilon'=0}} \left(r \rightarrow \infty\right) = G_{\mathrm{K}}^{\mathrm{\epsilon'=\epsilon}} - \frac{1}{2 \pi \beta \rho \mu^2} \frac{\left(\varepsilon - 1\right)^2}{\varepsilon} . \label{eq:GKinsulating}
\end{align}
}
Despite the differences discussed above, both $G_\mathrm{K}^{\epsilon'=\infty}$ and $G_\mathrm{K}^{\epsilon'=0}$ are intrinsically related to the true Kirkwood correlation factor $G_\mathrm{K}^{\epsilon'=\epsilon}$, as shown in Eq.~\ref{eq:GKmetallic} and ~\ref{eq:GKinsulating}. They differ from $G_\mathrm{K}^{\epsilon'=\epsilon}$ by a correction term arising from the integration of the dipolar correlation function $c_m^{\epsilon'}(r\rightarrow\infty)$, which accounts for the influence of the reaction field. These correction terms are constants that are independent of the simulation box volume $V$, as explicitly shown in Eq.~\ref{eq:GKmetallic} and ~\ref{eq:GKinsulating} for the metallic and insulating boundary conditions, respectively. Importantly, these expressions provide a practical means to compute Kirkwood correlation factors without requiring prior knowledge of water’s dielectric constant, which is necessary for imposing the KF boundary condition. They also imply that a converged value of $G_\mathrm{K}^{\epsilon'=\infty}$ or $G_\mathrm{K}^{\epsilon'=0}$ can only be obtained when the volume integral in Eq.~\ref{eq:KFeqn} is performed over the entire simulation box. Based on this framework, the true Kirkwood correlation factor GK can be extracted from $G_\mathrm{K}^{\epsilon'=\infty}$  or $G_\mathrm{K}^{\epsilon'=0}$  by subtracting the corresponding reaction field corrections. Using this approach, we obtain $G_\mathrm{K}^{\epsilon'=\epsilon}=2.32$ and $G_\mathrm{K}^{\epsilon'=\epsilon}=2.39$ from metallic and insulating boundary conditions, respectively, whose values are in close agreement with the Kirkwood correlation factor calculated directly under the KF boundary condition.

Similarly, an alternative approach to computing the Kirkwood correlation factor without prior knowledge of the dielectric constant was introduced by M.~Sprik \emph{et al.}~\cite{zhang2016computing}. In this method, the dipolar correlation function \( c_m^{\varepsilon'}\left(r\right) \) under the KF electric boundary condition is effectively reconstructed by combining results from two separate molecular dynamics simulations conducted under metallic and insulating boundary conditions. In doing so, the extra contributions from the reaction field in \( G_\mathrm{K}^{\epsilon'} \) are exactly canceled according to Eq.~\ref{eq:GKmetallic} and ~\ref{eq:GKinsulating}, following the relation
\begin{equation}
G_\mathrm{K}^{\epsilon'=\epsilon} = \frac{2 G_\mathrm{K}^{\epsilon'=\infty} + G_\mathrm{K}^{\epsilon'=0}}{3}. \label{eq:Combined}
\end{equation}
Using this combined boundary condition approach of Eq.~\ref{eq:Combined} in our simulations, we obtain a value of \( G_\mathrm{K}^{\epsilon'=\epsilon} = 2.44 \), which closely agrees with the values obtained from the other methods described above.

Finally, with knowledge of the Kirkwood correlation factor $G^{\epsilon'}_{\mathrm{K}}$ and the average dipole moment $\mu \sim 2.97$ Debyes, the dielectric constant of liquid water can be computed using Eq.~\ref{eq:KFeqn}. 
The value of $\mu$ is largely insensitive to the choice of electrical boundary conditions, reflecting its predominantly local character. Moreover, the molecular dipole moment $\mu$ exhibits rapid convergence in simulations, typically requiring only a few hundred picoseconds of sampling to obtain a well-converged estimate. 
In this regard, a consistent predicted dielectric constant $\varepsilon=102$ is reached under the metallic, KF, and the combined metallic–insulating boundary conditions as described in Fig.~\ref{fig:dipolecorrelation}(e). However, due to the equivalence between dipole correlation method and dipole fluctuation approach, the dielectric constant calculated under the insulating boundary condition is still subject to large relative errors, primarily because it requires inverting a small number approaching zero. As a result, achieving reliable convergence under the insulating boundary condition would require molecular dynamics trajectories extending to the microscend timescale, which exceeds the computational limits of our current simulations.

\section{CONCLUSIONS}
This study presents a comprehensive framework for computing the dielectric permittivity of liquid water using machine-learned potentials that incorporate long-range Coulombic interactions. By employing a dual deep neural network approach—combining the DPLR model and the Deep Wannier model, both trained on hybrid density functional theory data with the SCAN0 approximation—we accurately determine liquid water trajectories and its dielectric response under ambient conditions. The inclusion of long-range electrostatics, enabled by Deep Wannier predictions of the centers of maximally localized Wannier functions, ensures physically accurate modeling of correlated dipole fluctuations under different electric boundary conditions, which is essential for reliable evaluation of the dielectric permittivity.

Using linear response theory, we compute the dielectric constant through both the dipole fluctuation method and the Kirkwood dipole correlation formalism. Three electric boundary conditions—metallic, insulating, and Kirkwood-Fröhlich—are explicitly examined. From the dipole fluctuation method, we extract characteristic polarization relaxation times of approximately 0.3 $ps$ (longitudinal) and 8 $ps$ (transverse). We also introduce a unified approach to determine the Kirkwood correlation length ($\sim$ 16 \AA ) and Kirkwood correlation factor ($\sim 2.4$) based on the asymptotic behavior of dipole correlation functions across all boundary conditions. Consistently, our simulations yield a dielectric constant of approximately 102 from both computational approaches, irrespective of the boundary condition applied.

This methodology provides a robust and generalizable platform for modeling dielectric properties in polar liquids, offering key insights into the role of electrostatic boundary conditions and collective dipole dynamics. Future extensions may apply this framework to complex aqueous systems, such as electrolyte solutions ~\cite{Swartz2013} or confined water, to advance our understanding of dielectric behavior in chemical and biological contexts. The overestimation of the dielectric constant relative to experiment likely arises from two sources: (1) the absence of nuclear quantum effects (NQEs), which are known to soften the hydrogen-bond network due to proton delocalization ~\cite{zhang2021modeling, TangPRB2021} ; and (2) residual deficiencies in the exchange-correlation approximation within DFT, which tend to overestimate hydrogen bonding strength, leading to an over-structured liquid.

\begin{acknowledgments}
We thank Roberto Car and Linfeng Zhang for fruitful discussions. This work was supported by Seven Research, LLC. We also acknowledge partial support from the “Chemistry in Solution and at Interfaces” (CSI) Center funded by the U.S. Department of Energy through Award No. DE-SC0019394. This research used resources of the National Energy Research Scientific Computing Center (NERSC), which is supported by the U.S. Department of Energy (DOE), Office of Science under Contract No. DEAC02-05CH11231. This research includes calculations carried out on HPC resources supported in part by the National Science Foundation through major research instrumentation Grant No. 1625061 and by the U.S. Army Research Laboratory under Contract No. W911NF-16-2-0189. This research used resources of the Oak Ridge Leadership Computing Facility at the Oak Ridge National Laboratory, which is supported by the Office of Science of the U.S. Department of Energy under Contract No. DE-AC05-00OR22725.
\end{acknowledgments}

\section*{Data Availability}

The data that supports the findings of this study are available from the corresponding author upon reasonable request.

\bibliographystyle{aipnum4-2}
\bibliography{ref}
\clearpage
\end{document}